\documentstyle[floats,prb,aps,epsf]{revtex}
\begin{document}
\draft
\title{Surface electronic scattering in d-wave superconductors.}
\author{A.\ A.\ Golubov $^{1}$, M.\ Yu.\ Kupriyanov $^{2}$}
\address{$1)$ Department of Applied Physics, University of Twente,\\
P.O.Box 217 \\
7500 AE Enschede, The Netherlands\\
$2)$ Nuclear Physics Institute, Moscow State University, 119899 GSP \\
Moscow, Russia}
\maketitle

\begin{abstract}
Theoretical model of a rough surface in a d-wave superconductor is studied
for the general case of arbitrary strength of electron scattering by an
impurity layer covering the surface. Boundary conditions for quasiclassical
Eilenberger equations are derived at the interface between the impurity
layer and the d-wave superconductor. The model is applied to the
selfconsisrent calculation of the surface density of states and the critical
current in d-wave Josephson junctions.
\end{abstract}



Understanding the transport properties of surfaces and interfaces in high
temperature superconductors (HTS) is important for correct interpretation of
the experimental data and for applications of these materials. Up to now
surface peculiarities in $d$-wave superconductors were extensively discussed
in the framework of theoretical models based on quasiparticle reflection
from specular interfaces \cite{Hu,Tan1,Barash1,Rainer1}. The influence of
surface roughness in this approach was studied by the Ovchinnikov method 
\cite{Ovch}, namely, by introducing a thin impurity layer covering the
surface \cite{Barash1,Rainer1,Ovch,Rainer3}. The degree of roughness is then
measured by the ratio of the layer thickness $d$ to the quasiparticle mean
free path in the layer $\ell .$ This approach was used to study the smearing
of Andreev surface bound states by weak disorder $(d\ll \ell )$.

The regime of strong surface roughness, $d$ $\gg $ $\ell $, was discussed in 
\cite{GK98}. In this case the parameter-free boundary conditions were
derived at the interface between the superconductor and the disordered
layer. Similar conditions were also derived earlier in \cite{Kopnin} at
rough walls in superfluid $^3He-B$. As was shown in \cite{GK98}, an
isotropic order parameter is nucleated in the disordered layer even in the
absence of the bulk subdominant pairing interaction in the $s$-wave channel.
This leads to the formation of the gapless $s$-state at the interface. This
phenomenon was called an anomalous proximity effect between a $d$-wave
superconductor and a disordered layer.

In the present paper we consider the anomalous proximity effect between a $d$%
-wave superconductor and a surface impurity layer of thickness $d\ll \min
\left\{ \sqrt{\xi _0\ell },\ \xi _0\right\} $ in the general case of an
arbitrary electronic mean free path $\ell $.

We consider the interface oriented normally to the crystallographic $ab$
plane and assume elastic Born scattering. The normal and anomalous
quasiclassical propagators, $g$ and $f$ , depend on the coordinate $x$ in
the direction of the surface normal and on the angle $\theta $ between the
surface normal and quasiparticle trajectory and obey the quasiclassical
Eilenberger equations \cite{Eilenb}. Using the substitutions \cite{Schopohl} 
\begin{equation}
f=\frac{2a}{1+ab},\text{ }f^{+}=\frac{2b}{1+ab},\text{ }g=\frac{1-ab}{1+ab},
\label{Eq.1}
\end{equation}
we rewrite these equations in terms of the scalar differential equations of
the Ricatti type 
\begin{equation}
v\cos \theta \frac{db(x)}{dx}-\left[ 2\omega _n+\frac 1\tau \left\langle
g\right\rangle +b(x)(\Delta (x)+\frac 1{2\tau }\left\langle
f^{+}\right\rangle )\right] b(x)=-\Delta (x)-\frac 1{2\tau }\left\langle
f^{+}\right\rangle ,  \label{Eq.2}
\end{equation}
\begin{equation}
v\cos \theta \frac{da(x)}{dx}+\left[ 2\omega _n+\frac 1\tau \left\langle
g\right\rangle +a(x)(\Delta (x)+\frac 1{2\tau }\left\langle f\right\rangle
)\right] a(x)=\Delta (x)+\frac 1{2\tau }\left\langle f\right\rangle .
\label{Eq.3}
\end{equation}
The pair potential $\Delta $ should be found from the selfconsistency
equation 
\begin{equation}
\Delta \ln \frac T{T_c}+2\pi T\sum_{\omega >0}\left( \frac \Delta \omega
-\left\langle \lambda (\theta ,\theta ^{\prime })f\right\rangle \right) =0.
\label{Eq.4}
\end{equation}
Here $\omega _n=\pi T(2n+1)$ are the Matsubara frequencies, $v$ is the Fermi
velocity, $\tau =\ell /v$. Assuming a cylindrical Fermi surface, angle
averages are defined as $\left\langle ...\right\rangle =(1/2\pi
)\int_0^{2\pi }(...)d\theta $.

We consider pure $d$-wave interaction in the bulk with the coupling constant 
\cite{carbotte}

\[
\lambda (\theta ,\theta ^{\prime })\equiv \lambda _d(\theta ,\theta ^{\prime
})=2\lambda \cos (2(\theta -\alpha ))\cos (2(\theta ^{\prime }-\alpha )), 
\]
where $\alpha $ is the misorientation angle between the crystallographic $a$
axis and the surface normal. In this case the bulk anomalous propagator has
the $d$-wave symmetry

\begin{equation}
b(+\infty ,\theta )=a(+\infty ,\pi +\theta )=\frac{\sqrt{2}\Delta _\infty
\cos (2(\theta -\alpha ))}{\omega +\sqrt{\omega ^2+2\Delta _\infty ^2\cos
^2(2(\theta -\alpha ))}}.  \label{Eq.5}
\end{equation}
As will be shown below, this symmetry is violated near the interface.

Equations (\ref{Eq.2})-(\ref{Eq.4}) must be supplemented with the boundary
conditions connecting functions $b(0,\theta )$ and $a(0,\theta )$ which
describe respectively the electrons incident and reflected from the
interface. This can be done by matching the solutions in the impurity layer (%
$-d<x<0$) and in the clean d-wave region ($0<x$).

Since the thickness of the impurity layer $d\ll \xi _{eff}=\min \left\{ 
\sqrt{\xi _0\ell },\ \xi _0\right\} $, we can neglect the terms proportional
to $\omega $ and $\Delta $ in (\ref{Eq.2}) and (\ref{Eq.3}) and consider $%
\left\langle f\right\rangle $ and $\left\langle g\right\rangle $ as
spatially-independent constants,\ which must be determined selfconsistently.
Making use of the boundary condition \cite{Zai} at the totally reflecting
free specular interface $(x=-d)$ 
\begin{equation}
b(-d,-\theta )=a(-d,\theta ),  \label{Eq.8}
\end{equation}
we find the solution of the equations (\ref{Eq.2}), (\ref{Eq.3}) at $-d\leq
x\leq 0$ in the form 
\begin{equation}
\frac{Fb(x,-\theta )+G-1}{Fb(x,-\theta )+G+1}=\frac{Fb(0,-\theta )+G-1}{%
Fb(0,-\theta )+G+1}\exp \left\{ kx\right\} ,  \label{Eq.9}
\end{equation}
\begin{equation}
\frac{Fa(x,\theta )+G-1}{Fa(x,\theta )+G+1}=\exp \left\{ -k(d+x)\right\} 
\frac{Fb(-d,-\theta )+G-1}{Fb(-d,-\theta )+G+1}  \label{Eq.10}
\end{equation}
\[
k=\frac{\sqrt{\left\langle g\right\rangle ^2+\left\langle f\right\rangle ^2}%
}{\ell \cos (\theta )},\quad F=\frac{\left\langle f\right\rangle }{\sqrt{%
\left\langle g\right\rangle ^2+\left\langle f\right\rangle ^2}},\quad G=%
\frac{\left\langle g\right\rangle }{\sqrt{\left\langle g\right\rangle
^2+\left\langle f\right\rangle ^2}}. 
\]
From (\ref{Eq.9}), (\ref{Eq.10}) after a simple algebra we arrived at the
boundary condition at the interface between the clean $d-$wave
superconductor and disordered layer

\begin{equation}
a(0,\theta )=b(0,-\theta )\frac{(1-G\tanh (kd))}{(Fb(0,-\theta )+G)\tanh
(kd)+1}+F\frac{\tanh (kd)}{(Fb(0,-\theta )+G)\tanh (kd)+1}.  \label{Eq.11}
\end{equation}
Equation (\ref{Eq.11}) describes the fact that the amplitude of anomalous
Green's function along outgoing trajectories, $a(0,\theta )$, consists of
two parts. The first, specular part, is proportional to $b(0,-\theta )$ and
describes the correlation between the incoming $-\theta $ and reflected $%
\theta $ trajectories. The second, diffusive part, is proportional to $F$
and describes an average contribution of all the incident trajectories to
the outgoing one in the $\theta -$direction.

In the limit of weak disorder ($d\ll \ell )$ the boundary condition (\ref
{Eq.11}) reduces to the specular one (\ref{Eq.8}). At finite $d$ there is a
cone of angles $\arccos (d/\ell )\leq \theta \leq \pi /2$ in which the
scattering from the interface is rather diffusive than specular. The larger
is $d$ the smaller is the correlation between incoming and outgoing
trajectories.

In the limit of strong disorder $(d\gg \ell )$ the solutions (\ref{Eq.9})- (%
\ref{Eq.10}) transform to those of Usadel equations obtained in \cite{GK98}
and it follows from (\ref{Eq.11}) that 
\begin{equation}
a(0)=\frac{\left\langle f(0,\theta )\right\rangle }{1+\left\langle
g(0,\theta )\right\rangle }.  \label{Eq.12}
\end{equation}
This is exactly the condition used previously in \cite{GK98} for the
analysis of strongly disordered d-wave interfaces. According to this
condition the incoming and outgoing trajectories at the rough interface with
strong disorder are completely uncorrelated as a result of the scattering
from the impurity layer.

It is straightforward to find selfconsistent solutions of the set of
equations (\ref{Eq.2})-(\ref{Eq.5}), (\ref{Eq.11}). First, we have
numerically integrated the equation (\ref{Eq.2}) for $b(x)$ with the initial
condition (\ref{Eq.5}) along the trajectory from $x=+\infty $ (bulk) to $x=0$
(interface). Then the equation (\ref{Eq.3}) for $a(x)$ was integrated from $%
x=0$ to $x=+\infty $ with the initial condition (\ref{Eq.11}). The angle
averages $\left\langle f(0)\right\rangle $\ and $\left\langle
g(0)\right\rangle $ and the pair potential $\Delta (x)$\ were calculated
iteratively.

We have applied this model for calculation of the low-temperature
conductance of NID tunnel junction, which is expressed through the surface
density of states $N(0,\theta ;\varepsilon )$

\begin{equation}
R_NdI/dV=\frac{\int_{-\pi /2}^{\pi /2}N(0,\theta ;V)D(\theta )\cos \theta
d\theta }{\int_{-\pi /2}^{\pi /2}D(\theta )\cos \theta d\theta },
\label{Cond}
\end{equation}
where $R_N$ is the normal state junction resistance and $D(\theta )$ is the
barrier transmission coefficient. The angle-resolved surface density of
states is given by $N(0,\theta ;\varepsilon )=%
\mathop{\rm Re}
(g(0,\theta ;\omega =-i\varepsilon ))$.

Fig.1 shows the results of calculations for $\alpha =20^0$ and various
values of the surface scattering parameter $d/\ell $. We have chosen $\delta 
$-shaped potential barrier with the angular dependence of transmission $%
D(\theta )=D(0)\cos ^2\theta $. Zero-bias anomaly (ZBA), which is the
manifestation of the sign reversal of the d-wave pair potential, is strongly
broadened with increase of $d/\ell $. In the limit of weak disorder $d/\ell
\ll 1$ the results exactly correspond to those of Ref.\cite{Barash1}.

With the increase of $d/\ell $ ZBA is smeared out completely, while the
signatures of finite-energy Andreev bound states are still present. The
latter are due to quasiparticles trapped in the surface region with the
reduced pair potential $\Delta (x)$. As is seen from Fig.1, the
angle-averaged density of states in the disordered layer is gapless and has
a number of peaks at the energy below the bulk pair potential. Note that for
large values of $d/\ell $ there are several (weak) peaks corresponding to
more than one bound state, in contrast to the case of specular interface
when only a single bound state exists at finite energy \cite{Barash1}. This
is because the average trajectory length of trapped quasiparticles, which
determines the number of resonances, is larger than $\xi _0$ in the
diffusive case. Mathematically this fact is described by the boundary
condition Eq.\ref{Eq.11}.

Next we apply the model to the study the influence of surface roughness on
the Josephson supercurrent in a tunnel junction based on d-wave
superconductors (DID junction). The critical current is given by the
expression \cite{Zai,Bar}

\begin{equation}
eR_NI_c=\frac{\pi T\sum_{\omega >0}\int_{-\pi /2}^{\pi /2}[f_{1s}(\theta
)f_{2s}(\theta )-f_{1a}(\theta )f_{2a}(\theta )]D(\theta )\cos \theta
d\theta }{\int_{-\pi /2}^{\pi /2}D(\theta )\cos \theta d\theta }.
\label{IcRn}
\end{equation}
Here $f_{1,2s(a)}=(f_{1,2}\pm f_{1,2}^{+})/2$ and $f_{1,2}$ are the surface
Green's functions of the left (right) electrode.

The critical current was calculated numerically using the solutions for $%
f_{1,2}$ and assuming $D(\theta )=D(0)\cos ^2\theta $. We consider two
cases: (1) symmetric junction with the misorientation angles $\alpha
_1=\alpha _2=20^0$, (2) antisymmetric (mirror) junction with $\alpha
_1=-\alpha _2=20^0$. The results of calculations are shown in Fig.2. In
complete agreement with the results of Ref.\cite{Bar,Tan,Sam,Bag} the mirror
junction exhibits a nonmonotonous $I_c(T)$-dependence due to the anomaly $%
f_{1,2a}\sim 1/\omega _n$ at low frequencies related to the midgap surface
bound state. Like ZBA in conductance, this anomaly is smeared out with the
increase of the surface scattering parameter $d/\ell $.

In the strong scattering regime $d/\ell \gg 1$ the $I_cR_N$ product of the
DID junction becomes small and saturates at the level $eI_c^{SIS}R_N/2\pi
T_c\sim 10^{-3}$ (the corresponding number for a conventional SIS junction
is $eI_c^{SIS}R_N/2\pi T_c\simeq 0.44$). The anomalous contribution $%
f_{1a}f_{2a}$ vanishes in this case and $I_c$ for the symmetric and mirror
junctions coincide. However, as shown in inset in Fig.2, $I_c(T)$ becomes
nonmonotonous with an increase of $d/\ell $. The reason is the destructive
interference within the impurity layer: the phases of Eilenberger functions
on the incoming trajectories alternate due to the d-wave angular structure.
As a result, the angle average $\left\langle f_{1,2}\right\rangle $ vanishes
at $\omega _n\ll \pi T_c$, reaches a maximum at $\omega _n\sim \pi T_c$ and
falls down as $\omega _n^{-2}$ at $\omega _n\gg \pi T_c$. It is the property 
$f_{1,2s}$ $\rightarrow 0$ at small $\omega _n$ which is the reason of
decreasing $I_c(T)$ at low $T$. Variation of the misorientation angle $%
\alpha $ for $d/\ell \gg 1$ does not change the $I_c(T)$ behavior, while $%
I_cR_N$ product at $T=0$ scales as $\cos ^2(2\alpha )$.

In conclusion, we have derived the boundary conditions for the Eilenberger
functions at the rough surface of a d-wave superconductor and applied them
to the study of the crossover from specular to diffusive surface scattering.
The low-temperature conductance of NID tunnel junction and the Josephson
supercurrent in DID tunnel junction are calculated for arbitrary degree of
surface roughness. It is shown that the width of the anomalies in the
conductance and in the critical current is controlled by a single scattering
parameter $d/\ell $.

{\bf Acknowledgments}. This work was supported in part by INTAS 93-790ext
and by the Program for Russian-Dutch Research Cooperation (NWO).

\twocolumn

\begin{figure}
\par
\begin{center}
\mbox{\epsfxsize=0.9\hsize \epsffile{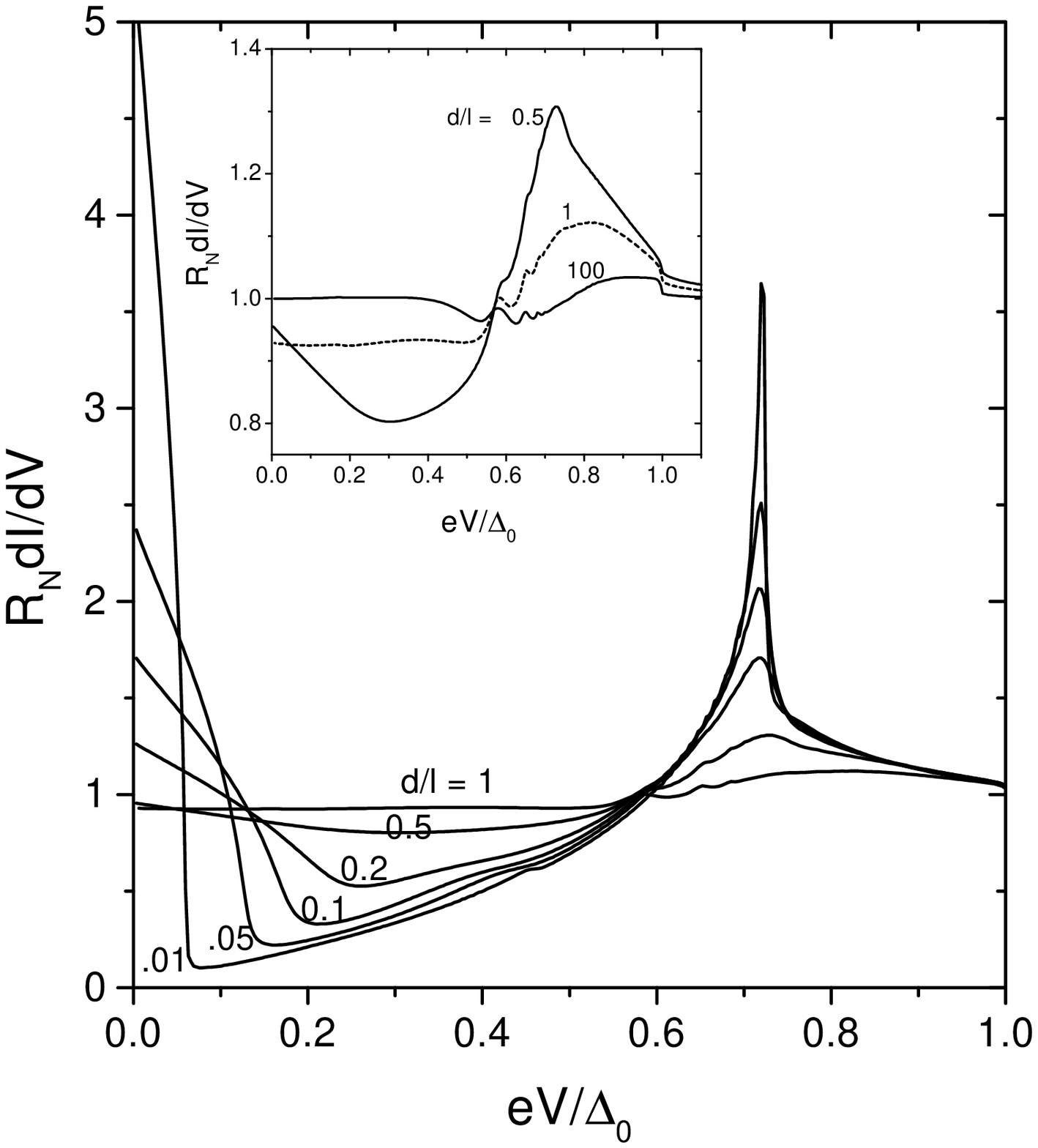}}
\end{center}
\caption{Smearing of ZBA in the low-temperature conductance of NID junction.
Inset demonstrates the oscillations for large $d/\ell $ due to the
finite-energy Andreev bound states.}
\end{figure}

\begin{figure}
\par
\begin{center}
\mbox{\epsfxsize=0.9\hsize \epsffile{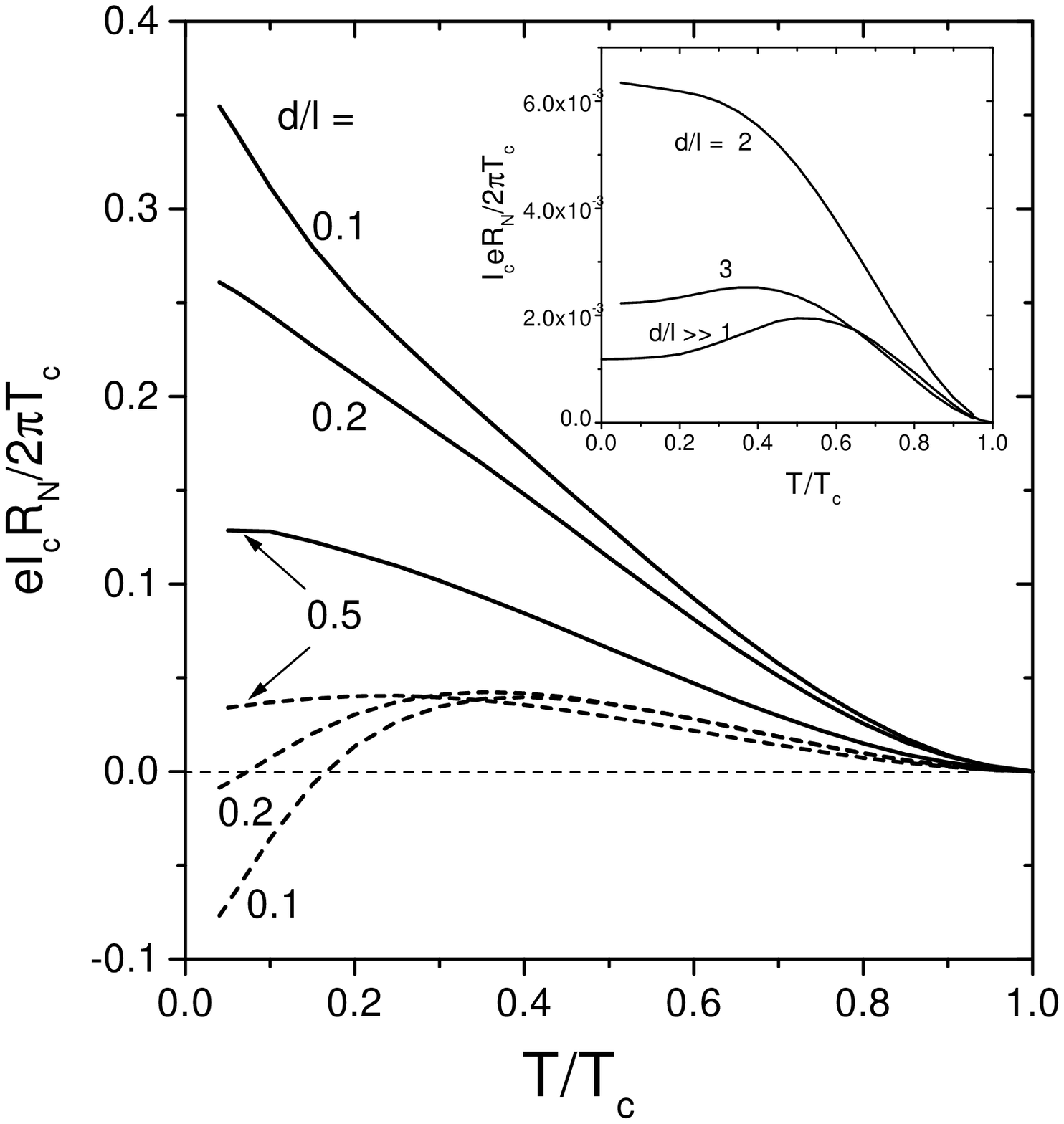}}
\end{center}
\caption{Influence of surface roughness on the temperature dependence of the
critical current in the DID tunnel junction. Dashed lines: symmetrical
junction $\alpha _1=\alpha _2=20^0.$ Dotted lines: mirror junction $\alpha
_1=-\alpha _2=20^0.$ Inset: nonmonotonous behavior of $I_c$ for $d/\ell \gg
1 $.}
\end{figure}

\end{document}